\documentclass[aps,prd,twocolumn,nofootinbib,preprintnumbers,superscriptaddress]{revtex4-1}
\usepackage[normalem]{ulem}
\usepackage{amsmath,amsfonts,graphicx,color,slashed}

\newcommand{\be}{\begin{equation}}
\newcommand{\ee}{\end{equation}}
\newcommand{\bea}{\begin{eqnarray}}
\newcommand{\eea}{\end{eqnarray}}

\begin{document}

\title{Pion crystals hosting topologically stable baryons}
\author{Fabrizio Canfora}
\email{canfora@cecs.cl}
\affiliation{Centro de Estudios Cient\'{\i}ficos (CECS), Casilla 1469,
Valdivia, Chile}

\author{Stefano Carignano}
\email{stefano.carignano@fqa.ub.edu}
\affiliation{Departament de F\'isica Qu\`antica i Astrof\'isica and Institut
de Ci\`encies del Cosmos, Universitat de Barcelona, Mart\'i i Franqu\`es 1,
08028 Barcelona, Catalonia, Spain.}

\author{Marcela Lagos}
\email{marcela.lagos@uach.cl}
\affiliation{Instituto de Ciencias F\'isicas y Matem\'aticas, Universidad
Austral de Chile,  Casilla 567, Valdivia, Chile}

\author{Massimo Mannarelli}
\email{massimo@lngs.infn.it}
\affiliation{INFN, Laboratori Nazionali del Gran Sasso, Via G. Acitelli,
22, I-67100 Assergi (AQ), Italy}

\author{Aldo Vera}
\email{aldo.vera@uach.cl}
\affiliation{Instituto de Ciencias F\'isicas y Matem\'aticas, Universidad
Austral de Chile,  Casilla 567, Valdivia, Chile}

\begin{abstract}
We construct analytic (3+1)-dimensional inhomogeneous and topologically
non-trivial pion systems using chiral perturbation theory. We discuss the
effect of isospin asymmetry with vanishing electromagnetic interactions as
well as some particular configurations with non-vanishing electromagnetic
interactions. The inhomogeneous configurations of the pion fields are
characterized by a non-vanishing topological charge that can be identified
with baryons surrounded by a cloud of pions. This system supports a
topologically protected persistent superflow. When the electromagnetic field
is turned on the superflow corresponds to an electromagnetic supercurrent.
\end{abstract}

\maketitle

\section{Introduction}

One of the main goals of the theoretical and experimental investigations in
Quantum Chromodynamics (QCD) is to determine the phases of hadronic matter
as a function of temperature, baryonic density and isospin asymmetry. In the
grand canonical ensemble this amounts to study the realization of the
hadronic phases as a function of the baryonic chemical potential, $\mu _{B}$%
, which encodes the baryonic density, and the isospin chemical potential, $%
\mu _{I}$, which determines the isospin asymmetry.

Since QCD is an asymptotically free theory we expect that at some large
energy scale hadrons melt liberating their quark and gluon content~\cite%
{Cabibbo:1975ig}. High temperature deconfined hadronic matter has been
realized in relativistic heavy-ion colliders, see for instance~\cite%
{Gyulassy:2004vg,Shuryak:2008eq,Satz:2012zza}, and it can possibly form in
the core of compact stars~\cite{Shapiro:1983du, Glendenning:1997wn}. In any
terrestrial heavy-ion experiment, as well as in the core of compact stars,
QCD is in the non-perturbative regime, posing a number of challenging
problems to the determination of the matter properties, see~\cite%
{Brambilla:2014jmp} for a review. In order to get insight on the properties
of hadronic matter, many different methods have been developed. At vanishing
baryonic density the deconfined phase can be studied by lattice QCD (LQCD)
methods~\cite{Smit:2002ug,Gattringer:2010zz}, but with increasing baryonic
density these numerical simulations become problematic: they are hampered by
the so-called sign problem, see~\cite{Muroya:2003qs, Schmidt:2006us,
deForcrand:2010ys, Philipsen:2012nu, Aarts:2015tyj} for recent progress in
this direction. For vanishing baryonic density and up to $\mu_I\simeq 2
m_\pi $, LQCD simulations are feasible~\cite{Alford:1998sd, Kogut:2002tm,
Kogut:2002zg,Kogut:2004zg,
Beane:2007es,Detmold:2008fn,Detmold:2008yn,Detmold:2011kw, Detmold:2012wc,
Endrodi:2014lja, Janssen:2015lda, Brandt:2016zdy, Brandt:2018omg,
Brandt:2017zck, Brandt:2018wkp} and their results can be compared with those
obtained by chiral perturbation theory ($\chi$PT)~\cite%
{Baym:1978sz,Kaplan:1986yq, Dominguez:1993kr,
Son:2000xc,Kogut:2001id,Birse:2001sn,Splittorff:2002xn,Loewe:2002tw,Loewe:2004mu,Loewe:2011tm,Mammarella:2015pxa,Adhikari:2015wva,Carignano:2016rvs, Loewe:2016wsk,Carignano:2016lxe,Adhikari:2018fwm,Lepori:2019vec,Adhikari:2019mdk,Tawfik:2019tkp,Mishustin:2019otg,Adhikari:2019zaj,Adhikari:2019mlf,Adhikari:2020exc,Adhikari:2020ufo,Agasian:2020plc, Adhikari:2020kdn}%
, or by Nambu--Jona-Lasinio (NJL) models~\cite%
{Barducci:1990sv,Toublan:2003tt, Barducci:2004tt,Barducci:2004nc, He:2005nk,
Ebert:2005cs, Ebert:2005wr, Mukherjee:2006hq, He:2005sp, He:2006tn,
Sun:2007fc, Andersen:2007qv,Abuki:2008tx, Abuki:2008wm,
Mu:2010zz,Xia:2013caa, Xia:2014bla, Chao:2018ejd,Khunjua:2019lbv,
Khunjua:2019nnv,Avancini:2019ego,Lu:2019diy,Cao:2019ctl,Khunjua:2020xws,
Khunjua:2020hbd, Mao:2020xvc}. In this way one can probe the robustness of
the obtained results. In particular, it is now well established that when
the isospin chemical potential exceeds the pion mass there is a second order
phase transition between the normal phase and the pion condensed phase, see~%
\cite{Mannarelli:2019hgn} for a recent review.

Various regions of the QCD phase diagram may be occupied by inhomogeneous
phases, see for instance \cite{Fukushima:2010bq,Anglani:2013gfu,Buballa:2014tba,Harada:2015lma,Kaplunovsky:2015zsa,Park:2019bmi}.
 The analysis of models in (1+1) and (3+1) dimensions has shown
that at low temperature some crystalline structures can be thermodynamically
stable: energetically favored with respect to the homogeneous phase. A quite
relevant result in this area has been the construction of exact crystalline
solutions of ordered solitons, see \cite{Basar:2008im, Basar:2008ki,
Basar:2009fg,Thies:2003br, Bringoltz:2009ym,Nickel:2009wj,
Karbstein:2006er,Takayama:1992eu,Schon:2000he}. Whether an ensemble of
charged pions may form an inhomogeneous Bose-Einstein condensate at
sufficiently low temperature is an interesting possibility~\cite%
{Gubina:2012wp,Andersen:2018osr}. An example of an inhomogeneous phase is
the chiral soliton lattice (CSL), which is an inhomogeneous pionic phase
supported by strong external fields~\cite{Brauner:2016pko, Huang:2017pqe}.
In these works the order parameter depends on only one spatial coordinate,
allowing in this way the use of tools developed in Gross-Neveu models~\cite%
{Gross:1974jv, Dashen:1975xh, Shei:1976mn, Feinberg:1996gz}. This fact
however prevents the condensate itself from having a non-trivial topological
charge.

Topological stability can be achieved in (3+1)-dimensional inhomogeneous
condensates. However, a detailed analysis of the electromagnetic
interactions of (3+1)-dimensional spatially modulated condensates is not
easy: in these situations the only available numerical results on crystals
of solitons treat the electromagnetic field as a fixed external field
neglecting the back reaction of the hadronic matter. It would be an
extremely useful result to achieve a sound analytic control on gauged
solitons with high topological charge and with crystalline order. Explicit
examples have been obtained either in lower dimensions or when some extra
symmetries (such as SUSY) are available (see \cite{Schroers:1995he,
Arthur:1996np, Gladikowski:1995sc, Cho:1995rw, Loginov:2018wbo,
Adam:2015qsa, Adam:2014xfa, Alonso-Izquierdo:2014cza, Chimento:2018elo}).

In the present paper we use zero temperature gauged two-flavor $\chi $PT to
construct an analytic (3+1)-dimensional pion inhomogeneous condensate
characterized by a non-vanishing topological charge. We achieve this result
by an appropriate choice of the condensate ansatz and of the gauge field
configuration. The presence of a topological charge prevents the decay of
the inhomogeneous condensate into an homogeneous phase, but it is not a
sufficient condition for stability. A classical argument by Landau and
Peierls is that in three or fewer dimensions the thermal fluctuations
destroy the condensates depending on only one spatial coordinate~\cite%
{Baym:1982ca}. This is the reason why we consider a (3+1)-dimensional
modulation, thus corresponding to a crystalline-like phase. Regarding the
stability of this kind of models, Skyrme and Derrick showed~\cite%
{Skyrme:1961vq, Skyrme:1961vr, Skyrme:1962vh, Derrick:1964ww} that they do
not support static solitonic solutions in flat, topologically trivial
(3+1)-dimensional space-time. We will circumvent this argument considering a
finite spatial volume, as finite volume effects together with non-trivial
boundary conditions at finite volume break the Derrick's scaling argument 
\cite{Derrick:1964ww}. These ways to avoid the Derrick no-go argument will
be combined using the generalized hedgehog-like ansatz introduced in \cite%
{Canfora:2013osa, Canfora:2013xja, PhysRevD.89.025007, Canfora:2014aia,
Canfora:2015xra, Ayon-Beato:2015eca, Tallarita:2017bks, Canfora:2017ivv,
Giacomini:2017xno, Astorino:2018dtr, Alvarez:2017cjm, Aviles:2017hro,
Canfora:2018clt, Canfora:2018rdz, Canfora:2019asc,Canfora:2019kzd,
Alvarez:2020zui, Canfora:2020kyj, Canfora:2020ppn}, that we will properly
extend at non-vanishing isospin chemical potential. In order to construct
topologically stable solitons the previous works~\cite{Canfora:2013osa,
Canfora:2013xja, PhysRevD.89.025007, Canfora:2014aia, Canfora:2015xra,
Ayon-Beato:2015eca, Tallarita:2017bks, Canfora:2017ivv, Giacomini:2017xno,
Astorino:2018dtr, Alvarez:2017cjm, Aviles:2017hro, Canfora:2018clt,
Canfora:2018rdz, Canfora:2019asc,Canfora:2019kzd, Alvarez:2020zui,
Canfora:2020kyj, Canfora:2020ppn} needed to consider a time dependent
modulation with time dependent boundary conditions. Within the present work
we show that for non-vanishing isospin chemical potential, it is possible to
obtain a topologically stable crystalline phase with static background field
and with time independent boundary conditions. This implies that the
proposed crystalline phase may be studied in LQCD simulations, which employ
static boundary conditions.

Regarding the use of $\chi $PT, we remark that it is quantitatively under
control for $\mu _{I}<\Lambda _{c}\sim 1$ GeV, corresponding to the critical
scale of $\chi $PT. This effective field theory is based on two key
ingredients: the global symmetries of QCD and an appropriate low momentum
expansion~\cite{Weinberg:1978kz, Gasser:1983yg, Georgi:1985kw,
Leutwyler:1993iq, Ecker:1994gg, Leutwyler:1996er, Pich:1998xt,
Scherer:2002tk, Scherer:2005ri}. The results obtained within $\chi $PT agree
with those of other methods for $\mu_I \le 2 m_\pi$, see the discussion in~%
\cite{Mannarelli:2019hgn}, corroborating the reliability of this effective
field theory. Remarkably, $\chi $PT can also be used to study a variety of
gauge theories with isospin asymmetry, including two color QCD with
different flavors~\cite{Kogut:1999iv,Kogut:2000ek,
Hands:2000ei,Kogut:2001na, Brauner:2006dv,Braguta:2016cpw, Adhikari:2018kzh}.

It is worth to emphasize that, while we shall study the inhomogeneous
condensates in the context of $\chi $PT, the results reported in~\cite%
{Alvarez:2020zui, Canfora:2020kyj, Canfora:2020ppn} strongly suggest that
the existence of such condensates (as well as their explicit functional
forms) are very robust. In particular, the construction is not spoiled
neither by subleading corrections in the 't Hooft expansion nor by replacing
the $SU(2)$ internal symmetry with an $SU(N) $ internal symmetry group.
Hence, it is very natural to think that the results obtained in the present
manuscript could be valid even beyond $\chi $PT: we hope to come back on
this very interesting issue in a future publication.

This paper is organized as follows. In Sec.~\ref{sec:chipt} we briefly
review the application of $\chi$PT to meson condensation. In Sec.~\ref%
{sec:inhomogeneous} we discuss the inhomogeneous pion phase for vanishing
electromagnetic fields. In Sec.~\ref{sec:gauge} we consider the gauged model
with a particular configuration of the gauge fields. We draw our conclusions
in Sec.~\ref{sec:conclusions}.

We use the Minkowski metric $\eta_{\mu\nu} =(1,-1,-1,-1)$ and natural units $%
c= \hbar =1$.


\section{The $\protect\chi$PT description of meson condensation}

\label{sec:chipt} 

The low-energy properties of pions can be described by $\chi$PT~\cite%
{Weinberg:1978kz, Gasser:1983yg, Georgi:1985kw, Leutwyler:1993iq,
Ecker:1994gg, Leutwyler:1996er, Pich:1998xt, Scherer:2002tk, Scherer:2005ri}%
, which is grounded on the global symmetries of QCD and uses an expansion in
exchanged momenta. In this approach the pion fields can be collected in the
unimodular field 
\begin{equation}  \label{eq:Sigma}
\Sigma = e^{i \bm \alpha \cdot \bm \sigma} =\mathbf{1}_{2} \cos \alpha \pm i
N \sin \alpha\,,
\end{equation}
where $N=\bm n \cdot \bm \sigma$, with $\bm \sigma$ the Pauli matrices, and $%
\bm \alpha =\alpha \bm n $; we shall call $\alpha$ the radial field while $%
\bm n$ is a unimodular field in isospin space. This is a convenient
representation because it allows us to simplify the calculations; we shall
see below how these fields are related to the standard pion fields. The
leading order $SU(2)$ $\chi$PT Lagrangian including the electromagnetic
interaction can be written as 
\begin{align}
\mathcal{L}& =\frac{f_{\pi }^{2}}{4} \mathrm{Tr}\left[- \left( \Sigma ^{\mu
}\Sigma _{\mu }\right) + m_{\pi }^{2}\left( \Sigma +\Sigma ^\dagger\right) %
\right] -\frac{1}{4}F_{\mu \nu }F^{\mu \nu } \,,  \label{eq:lagrangian}
\end{align}
where the pion decay constant, $f_{\pi } \simeq 93 $ MeV, and the assumed
degenerate pion masses, $m_{\pi } \simeq 135$ MeV, are phenomenological
constants. Also 
\begin{align}  \label{eq:sigmamu}
\Sigma _{\mu } =\Sigma^{-1}D_{\mu }\Sigma = \Sigma _{\mu }^{j}\sigma_{j} \ .
\end{align}
The field strength is $F_{\mu \nu }=\partial _{\mu }A_{\nu }-\partial _{\nu
}A_{\mu }$ where $A^\mu $ is the electromagnetic gauge field and the
covariant derivative is defined as 
\begin{equation}
D_{\mu }\Sigma=\partial_{\mu }\Sigma+i \widetilde{A}_{\mu }\left[
\sigma_{3},\Sigma\right] \ ,
\end{equation}%
where $\partial _{\mu }$ is the usual partial derivative and 
\begin{equation}
\widetilde{A}^{\mu }=\left( \frac{\mu _{I}}{2}+A^{0},\bm{A}\right) \ ,
\label{eq:deftilda}
\end{equation}%
where we have included the effect of the isospin chemical potential, $\mu_I$%
. Regarding the gauge field potential, we shall assume that it is
self-consistently generated by the pion distribution. The classical field
equations read 
\begin{gather}
D_{\mu }\Sigma ^{\mu }+\frac{m_\pi^{2}}{2}\left( \Sigma
-\Sigma^\dagger\right) = 0\ ,  \label{eq:NLSM} \\
\partial_{\mu }F^{\mu \nu } = J^{\nu }\ ,  \label{eq:maxwellNLSM}
\end{gather}
where 
\begin{equation}
J^{\mu }=i \frac{f_\pi^2}{2}\text{Tr}\left[ \Sigma^{\mu}
(\Sigma^\dagger\sigma_{3}\Sigma-\sigma_{3}) \right] \,,  \label{maxcurrent}
\end{equation}
is the pion current generated by the electromagnetic field.

To make contact with the usual pion representation we expand Eq.~%
\eqref{eq:Sigma} retaining the leading order in $\bm \alpha$. In this way
Eq.~\eqref{eq:NLSM} yields 
\begin{equation}
\left( D_\mu D^\mu + m_\pi^2\right) \bm \alpha =0\,,
\end{equation}
which is the Klein-Gordon equation for three scalar fields. This equation
can be diagonalized to 
\begin{align}
(\partial_\mu \partial^\mu + m_\pi^2) \alpha_3&=0 \ , \\
(\partial_\mu \partial^\mu + 4 i \tilde A^\mu \partial_\mu + m_\pi^2 - 4
\tilde A^\mu \tilde A_\mu ) (\alpha_1 + i \alpha_2) &=0 \ , \\
(\partial_\mu \partial^\mu - 4 i \tilde A^\mu \partial_\mu + m_\pi^2 - 4
\tilde A^\mu \tilde A_\mu ) (\alpha_1 - i \alpha_2) &=0 \ ,
\end{align}
which explicitly show that $\alpha_3$ is the neutral field, while $\pi_\pm
\propto (\alpha_1 \pm i \alpha_2)$ correspond to the two charged pion
fields. For vanishing electromagnetic potential, the charged scalar fields
have dispersion law 
\begin{equation}  \label{eq:dispersion_normal}
E_\pm = \pm \mu_I \pm \sqrt{\bm p^2 + m_\pi^2} \, ,
\end{equation}
which manifestly shows that for $|\mu_I|= m_\pi$ a massless mode appears,
signaling a transition to the homogeneous pion condensed phase.

\subsection{The homogeneous phase}

Let us briefly recall the most important results of the two-flavor
homogeneous and time independent pion condensed phase. This phase is
characterized by the condensation of one of the two charged pion fields,
which induces the spontaneous symmetry breaking 
\begin{equation}
\underbrace{U(1)_I\times U(1)_Y}_{\displaystyle\supset U(1)_Q} \times U(1)
_B \to \underbrace{U(1)_Y\times U(1)_B}_{\displaystyle\not\supset U(1)_Q}\,,
\end{equation}
where $U(1)_I$, $U(1)_Y$ and $U(1)_B$ are three unitary groups associated
with the third component of isospin, hypercharge and baryonic number,
respectively, see~\cite{Mannarelli:2019hgn}. Whereas $U(1)_Q$ is the gauge
group of the electromagnetic interaction. The condensation happens at $%
|\mu_I| = m_\pi$ where one of the pion modes becomes massless, see Eq.~%
\eqref{eq:dispersion_normal}. This mode corresponds to the Nambu-Goldston
boson (NGB) associated with the spontaneous $U(1)_I$ breaking. Neglecting
the electromagnetic interaction the broken phase is a superfluid, while it
is an electromagnetic superconductor if the symmetry group is gauged.

In $\chi$PT the pion condensate can be introduced assuming that the
unimodular field $\Sigma$ in Eq.~\eqref{eq:Sigma} takes a non-trivial vev, $%
\bar \Sigma$, which can be determined treating $\alpha$ and $N$ as
variational parameters. The most general ansatz for the unit vector
background field is 
\begin{align}
n^{1}& =\sin \Theta \cos \Phi \ ,\ \ n^{2}=\sin \Theta \sin \Phi \ ,\ \
n^{3}=\cos \Theta \ ,  \label{eq:unitvector}
\end{align}
where $\Theta$ and $\Phi$ are two variational angles.

Upon substituting this vev in Eq.~\eqref{eq:lagrangian} one obtains~\cite%
{Mammarella:2015pxa} 
\begin{equation}  \label{eq:L0_tilde}
\mathcal{L}_\text{hom} =f_\pi ^2 m_\pi^2 \cos\alpha +2 {f_\pi^2}
\sin^2\alpha\tilde A^\mu \tilde A_{\mu}\sin^2\Theta - \frac{1}{4}
F_{\mu\nu}F^{\mu\nu}\,,
\end{equation}
which has the following well known feature: the normal phase is stable only
for $|\mu_I|< m_\pi$. In this case $\cos {\alpha}=1$, thus $\bar \Sigma=%
\text{diag}(1,1)$, while $\Theta$ and $\Phi$ are undetermined. The
associated vacuum pressure and energy density are respectively given by 
\begin{align}
p_\text{N}=f_{\pi }^{2}m_{\pi }^{2}\,, \qquad \epsilon_\text{N} = - p_\text{N%
}\,,  \label{eq:vacuum}
\end{align}
while the number density is zero.

When $|\mu_I|> m_\pi$ the system makes a second order phase transition to
the homogeneous {\ charged} pion condensed phase. In this case the trivial
vacuum is unstable and the energetically favored phase is characterized by 
\begin{equation}  \label{eq:alpha_hom}
\cos\alpha_0 = \frac{m_\pi^2}{\mu_I^2}\,,\qquad \Theta_0=\frac{\pi}{2}\,,
\end{equation}
while $\Phi$ can take any arbitrary value. Since the static Lagrangian in
Eq.~\eqref{eq:L0_tilde} does not depend on $\Phi$, the potential has a flat
direction orthogonal to the $3$-direction in isospin space which is spanned
by the NGB. In the broken phase, the normalized pressure and the energy
density (obtained subtracting the vacuum values), are respectively given by 
\begin{align}  \label{eq:phom}
p&=\frac{f_\pi^2}{2 \mu_I^2} \left(\mu_I^2-m_\pi^2 \right)^2\,, \\
\epsilon&= \frac{f_\pi^2}{2 \mu_I^2}(\mu_I^2-m_\pi^2)(\mu_I^2+3m_\pi^2) \ ,
\label{eq:epsilonhom}
\end{align}
which are positive and vanish at the phase transition point.

The aspects of the homogeneous phase that will be relevant in the discussion
of the inhomogeneous phases are the followings: In the broken phase the
value of the radial angle $\alpha$ depends on $\mu_I^2$ by Eq.~%
\eqref{eq:alpha_hom}, while the normal phase is characterized by $\alpha = 2
k\pi$, with $k$ an integer. Restricting to $k=1$, $\alpha$ can only assume
values in the intervals $[0,\pi/2]$ and $[3\pi/2,2\pi]$; values outside this
intervals cannot be attained in the homogeneous phase. The angle $\Theta$ is
not specified in the unbroken phase, but equals $\pi/2$ in the broken phase.
The flat direction of the potential corresponds to the one orthogonal to $%
n_3 $ and spanned by the angle $\Phi$. Finally, in the broken phase any
electromagnetic field is screened, as indicated by the second term on the
rhs of Eq.~\eqref{eq:L0_tilde}, meaning that supercurrents can circulate
with vanishing resistance.


\section{The inhomogeneous topological phases for vanishing gauge fields}

\label{sec:inhomogeneous} 
We begin with studying the inhomogeneous phases with vanishing gauge fields.
For the appearance of the inhomogeneous topological phases, finite volume
effects are of crucial importance. We take them into account using the
following metric 
\begin{equation}
ds^{2}=dt^{2}- \ell^2\left( dr^{2}+d\theta ^{2}+d\phi ^{2}\right) \ ,
\label{Minkowski}
\end{equation}%
where 
\begin{equation}  \label{eq:ell}
\ell=\frac{b}{m_\pi} \,,
\end{equation}
with $b$ a real number, is the typical dimension of the system. With this
coordinate choice the derivative operator turns to $\partial_\mu = (\frac{%
\partial}{\partial t}, \frac{1}{\ell} \frac{\partial}{\partial r},\frac{1}{%
\ell} \frac{\partial}{\partial \theta},\frac{1}{\ell} \frac{\partial}{%
\partial \phi} )$. The adimensional coordinates $r$, $\theta $ and $\phi $
have ranges 
\begin{equation}
0\leq r\leq 2\pi \ ,\quad 0\leq \theta \leq \pi \ ,\quad 0\leq \phi \leq
2\pi \,,  \label{eq:ranges}
\end{equation}%
meaning that we are considering pions in a cell of volume $4\pi ^{3} \ell^3$%
. For the ground state solution we assume the unimodular form of Eq.~%
\eqref{eq:Sigma} that is now promoted to be a classical field, meaning that $%
\alpha \equiv \alpha(x^\mu)$, $\Theta \equiv \Theta(x^\mu) $ and $\Phi\equiv
\Phi(x^\mu)$ with appropriate Dirichlet boundary conditions. In particular,
we demand that 
\begin{equation}  \label{eq:BC1}
n(t,r,0,\phi) = - n(t,r,\pi,\phi)\,, \quad n(t,r,\theta,0) = n(t,r,\theta, 2
\pi) \ ,
\end{equation}
and that 
\begin{equation}  \label{eq:BC2}
\Sigma(t,0,\theta,\phi) = \pm \Sigma(t,2\pi,\theta,\phi)\,.
\end{equation}
As we will see, these boundary conditions allow us to have a non-vanishing
topological charge. Different boundary conditions can be accordingly
implemented.

For vanishing electromagnetic fields, the matter effective Lagrangian in Eq.~%
\eqref{eq:lagrangian} can be rewritten as 
\begin{align}  \label{eq:Lgen}
\mathcal{L}_\text{m} = & \frac{f_\pi^2}2 \left[ \partial_\mu \alpha
\partial^\mu \alpha +\sin^2{\alpha} \partial_\mu \Theta \partial^\mu \Theta
+ 2 m_\pi^2 \cos\alpha \right.  \notag \\
&\left. +\sin^2{\alpha} \sin^2{\Theta}(\partial_\mu \Phi -\mu_I \delta_{\mu
0})(\partial^\mu \Phi-\mu_I \delta^{\mu 0}) \right]\,,
\end{align}
showing that the three classical fields $\alpha, \Theta$ and $\Phi$ are
non-linearly interacting. Solving the classical problem amounts to find a
solution of Eqs.~\eqref{eq:NLSM} and \eqref{eq:maxwellNLSM} with vanishing
gauge field, which is equivalent to solve the three coupled differential
equations 
\begin{align}
\partial_\mu\partial^\mu\Phi =& - (\partial_\mu\Phi - \mu_I \delta_{\mu_0})
\partial^\mu (\log(\sin^2\alpha \sin^2\Theta) )\,,  \label{eq:Phi} \\
\partial_\mu\partial^\mu\Theta =& -2 \cot
\alpha\,\partial^\mu\Theta\partial_\mu\alpha+ \frac{\sin2\Theta}{2} K \,,
\label{eq:Theta} \\
\partial_\mu \partial^\mu \alpha = & - m_\pi^2 \sin\alpha + \frac{\sin(2
\alpha)}{2}( \partial_\mu \Theta \partial^\mu \Theta + K \sin^2\Theta )\,,
\label{eq:alpha}
\end{align}
where $K= (\partial_\mu\Phi - \mu_I \delta_{\mu_0})(\partial^\mu\Phi - \mu_I
\delta^{\mu_0})$, which is a non-trivial task. From the discussion of the
homogeneous phase, we expect that the $\Phi$ field corresponds to a NGB.
Indeed, in the above equations $\Phi$ is the only field that is massless and
with derivative interactions, as appropriate for a NGB. This is more evident
assuming that $\Phi \equiv \Phi(t,\phi)$ and that $\alpha$ and $\Theta$
depend only on $r$ and $\theta$. Then Eq.~\eqref{eq:Phi} simplifies to the
free field equation of a massless scalar field 
\begin{align}  \label{eq:KG_Phi}
\partial_\mu\partial^\mu\Phi = \left(\frac{\partial^2}{\partial t^2} -\frac{1%
}{\ell^2}\frac{\partial^2}{\partial \phi^2} \right) \Phi= 0\,,
\end{align}
which has the standard free field propagating solution. However, we are
interested in solitonic solutions, therefore we shall consider the solution
of Eq.~\eqref{eq:KG_Phi} that depend linearly on $t$ and $\phi$ of the form 
\begin{equation}
\Phi =\frac{a }{\ell}t-p \phi + \Phi_0 \,,
\end{equation}
where $a$ and $\Phi_0 $ are real numbers and $p\in \mathbb{Z}$. In this way
Eqs.~\eqref{eq:Theta} and \eqref{eq:alpha} yield 
\begin{align}
\nabla^2 \Theta =&- 2 \cot \alpha\,(\partial_r\Theta\partial_r\alpha
+\partial_\theta\Theta\partial_\theta\alpha ) -K \ell^2 \frac{\sin2\Theta}{2}%
\, \,,  \label{eq:Theta2} \\
\nabla^2 \alpha = & \frac{1}{2}\sin(2 \alpha)\left((\partial_\theta
\Theta)^2+(\partial_r \Theta)^2 - K\ell^2 \sin^2(\Theta) \right)  \notag \\
& + m_\pi^2\ell^2 \sin\alpha \,,  \label{eq:alpha2}
\end{align}
where $\nabla^2 =(\frac{\partial^2}{\partial \theta^2} + \frac{\partial^2}{%
\partial r^2}) $ and 
\begin{equation}
K=\left(\frac{a}{\ell} -\mu_I\right)^2-\frac{p^2}{\ell^2}\,,
\end{equation}
is now a constant. This system of equations is still complicated, however we
can obtain analytical solutions in some particular cases.

If $\alpha$ is a constant then Eq.~\eqref{eq:Theta2} becomes a
sine-Gordon-like equation with solution 
\begin{equation}  \label{eq:sineG_theta}
\Theta(\bar \theta, \bar r ) = 2 \arctan\left[ \frac{\sinh(w \bar \theta/%
\sqrt{w^2-1} )}{w \cos(\bar r/\sqrt{w^2-1})} +\delta \right]\,,
\end{equation}
where $\bar r = \ell r K/2 $ and $\bar \theta = \ell \theta K/2 $, while $w$
and $\delta$ are two constants that depend on the boundary conditions.
Asking that $\Theta(0,\bar r) =0$, we readily fix $\delta=0$. Then,
according with Eq.~\eqref{eq:BC1}, we should demand that for any $\bar r$, $%
\Theta(0,\bar r) =(2 k+1) \pi$, with $k$ integer. However, this is not
compatible with Eq.~\eqref{eq:sineG_theta}. Moreover, the only solution with 
$\alpha$ constant of Eq.~\eqref{eq:alpha2} is $\alpha= n \pi$ with $n$
integer, corresponding to the homogeneous normal phase. For these reasons we
shall not consider this solution anymore.

\begin{figure}[h!]
\includegraphics[width=0.45\textwidth]{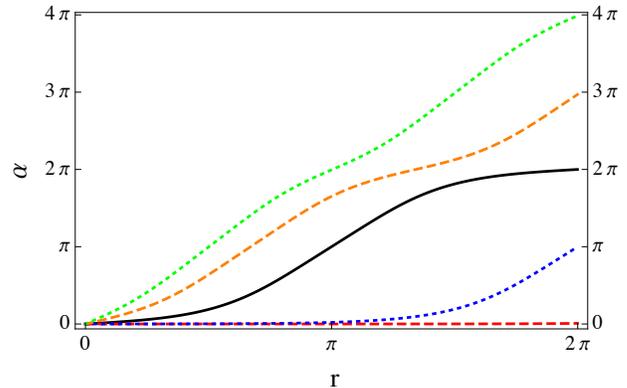}
\caption{Modulation of the radial mode, $\protect\alpha(r)$, obtained
numerically solving Eq.~\eqref{eq:diffalpha} for $q=1$ and $\ell=1/m_\protect%
\pi$ and assuming $\protect\alpha(0)=0$. We report the results obtained with
five different values of the integration constant $\protect\eta_0$, see Eq.~ 
\eqref{eq:eta}, to match the right boundary condition $\protect\alpha(2%
\protect\pi)= n \protect\pi$. For $\protect\eta_0=0$, red dashed line, the
radial mode identically vanishes. With non-vanishing values of $\protect\eta%
_0$ the radial field monotonically grows and the right boundary value $%
\protect\alpha(2\protect\pi)$ increases with increasing values of $\protect%
\eta_0$. }
\label{fig:alpha}
\end{figure}

\begin{figure}[h!]
\includegraphics[width=0.45\textwidth]{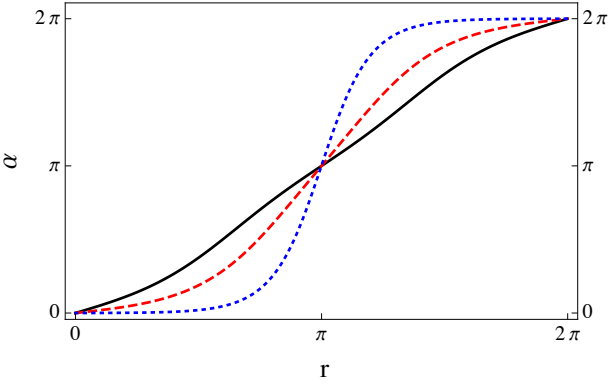}
\caption{Modulation of the radial mode, $\protect\alpha(r)$, obtained
numerically solving Eq.~\eqref{eq:diffalpha} for three different values of
the lattice size, see Eq.~\eqref{eq:ell}. The solid black line corresponds
to $b=0.5$, the dashed red line to $b=1$ and the dotted blue line to $b=2.5$%
. }
\label{fig:alpha_b}
\end{figure}

A different class of solutions can be obtained considering $\Theta\equiv
\Theta(\theta)$ and $\alpha \equiv \alpha(r)$. In order to make Eq.~%
\eqref{eq:alpha2} independent of $\theta$ we have to assume that $\Theta$
depends linearly on $\theta$ and that $K=0$. It follows that in this case,
the unit vector in Eq.~\eqref{eq:unitvector} is modulated as 
\begin{equation}
\Phi =\frac{a}\ell t-p \phi + \Phi_0 \ ,\ \Theta =q \theta + \Theta_0 \,,
\label{eq:general}
\end{equation}
where $a, \Theta_0$ and $\Phi_0 $ are real numbers, $p,q \in \mathbb{Z}$
with $q$ odd. As remarked in~\cite{Canfora:2013osa, Canfora:2013xja,
PhysRevD.89.025007, Canfora:2014aia, Canfora:2015xra, Ayon-Beato:2015eca,
Tallarita:2017bks, Canfora:2017ivv, Giacomini:2017xno, Astorino:2018dtr,
Alvarez:2017cjm, Aviles:2017hro, Canfora:2018clt, Canfora:2018rdz,
Canfora:2019asc,Canfora:2019kzd, Alvarez:2020zui, Canfora:2020kyj,
Canfora:2020ppn}, the time dependence of the unit vector is sufficient to
avoid the Derrick's no-go theorem on the existence of solitons in non-linear
scalar field theories. It corresponds to a unit vector rotating at constant
speed around the $3$-direction in isospin space, which is precisely the flat
potential direction discussed in the homogenous phase. However, this is not
a necessary condition, indeed the considered system is confined in a finite
volume and this suffices to avoid the scaling argument of the Derrick's
theorem. It remains to impose the condition $K=0$, which can be written as 
\begin{equation}  \label{eq:a}
a= \ell \mu_I + p\,,
\end{equation}
where $p$ can be a positive or a negative integer. This condition relates
the parameter of the classical field $\Phi$ with the isospin chemical
potential. Remarkably, it is possible eliminate any time dependence:
imposing that $a=0$ it follows that 
\begin{equation}  \label{eq:simplification}
p= - \mu_I \ell \,,
\end{equation}
which relates in a clear way the finite volume size and the isospin chemical
potential. Notice that it is possible to eliminate the time dependence only
for non-vanishing isospin chemical potentials. The advantage of this choice
is that the boundary condition at $\phi=0$ and $\phi= 2\pi$ become time
independent and can be possibly implemented in LQCD simulations. Finally,
the $\theta$ dependence of the $\Theta$ field allows us to span all its
possible values, including the one in Eq.\eqref{eq:alpha_hom}, corresponding
to the maximum of the homogeneous Lagrangian. Upon substituting Eq.~%
\eqref{eq:general} in the differential equation~\eqref{eq:alpha} one readily
finds that it can be written as~\cite{Canfora:2019kzd} 
\begin{align}
\frac{\partial^2 \alpha}{\partial r^2} = & m_\pi^2 \ell^2 \sin\alpha + \frac{%
q^2}{2} \sin(2 \alpha)\,,  \label{eq:alpha_sol}
\end{align}
meaning that the modulation of the $\alpha$ field does not explicitly depend
on the isospin chemical potential. This seems at odds with the result of the
homogeneous broken phase in Eq.~\eqref{eq:alpha_hom}. However, $\ell$ and $%
\mu_I$ are related by Eq.~\eqref{eq:simplification}, therefore there is an
implicit dependence on $\mu_I$. It is indeed possible to obtain the
homogeneous solution from Eq.~\eqref{eq:alpha_sol} noticing that in this
case it gives 
\begin{equation}
\cos \bar\alpha = -\frac{m_\pi^2}{\mu_I^2} \left(\frac{p}{q}\right)^2\,,
\end{equation}
which is indeed similar to Eq.~\eqref{eq:alpha_hom}. Therefore, the
homogeneous phase corresponds to the prescription 
\begin{equation}  \label{eq:cond_om}
p=\pm q\,, \qquad \alpha_0 = \bar\alpha +\pi\,.
\end{equation}
Regarding the general $\alpha(r)$ modulation, the radial field second order
differential equation \eqref{eq:alpha_sol} can be recast as the first order
differential equation 
\begin{equation}
\frac{\partial\alpha}{\partial r} = \eta (\alpha) \,,  \label{eq:diffalpha}
\end{equation}
where we can determine the function $\eta (\alpha)$ noticing that 
\begin{equation}
\frac{\partial^2\alpha}{\partial r^2} =\frac{\partial\eta}{\partial r} =%
\frac{1}{2}\frac{\partial\eta^2}{\partial \alpha} \,,
\end{equation}
and then from Eq.~\eqref{eq:alpha_sol} we obtain 
\begin{equation}
\eta (\alpha) =\pm \sqrt{\eta_0^2+ 2 m_\pi^2 \ell^2 (1-\cos{\alpha}) +q^2
\sin^2(\alpha)}\,,  \label{eq:eta}
\end{equation}
where $\eta_0$ is an adimensional integration constant and the positive
(negative) sign corresponds to solutions with increasing (decreasing) values
of $\alpha(r)$ in the interval $r \in [0,2\pi]$. In the following we shall
assume that $\eta$ is non-negative and that $\eta_0$ is such that for a
given $q$, 
\begin{equation}
\int_{0}^{n\pi }\frac{d\alpha}{\eta (\alpha)} =2\pi \,,
\label{eq:intconstcond}
\end{equation}%
where we have assumed the boundary conditions 
\begin{equation}  \label{eq:BCalpha}
\alpha(0) =0 \qquad \text{and} \qquad \alpha(2 \pi) = n \pi\,,
\end{equation}
where $n$ is an integer. Even (odd) values of $n$ correspond to periodic
(antiperiodic) boundary condition in the $r$ direction, see Eq.~%
\eqref{eq:BC2}. The above integral can be evaluated in terms of elliptic
functions; alternatively, one can fix $\eta_0$ by 
\begin{equation}
n \pi= \alpha(2\pi) = \int_{0}^{2\pi} dr \ \eta(\alpha(r))\,,
\end{equation}
which follows from Eq.~\eqref{eq:diffalpha}. Therefore, for a given value of 
$q$ and $\ell$, the integration constant $\eta_0$, determines the value of
the $\alpha$ field at the right boundary, which, as we shall see, is linked
to the topological charge~\cite{Callan:1983nx,Piette:1997ny}. We report in
Fig.~\ref{fig:alpha} the plot of the radial field as a function of $r $ for $%
q=1$ and $\ell=1/m_\pi$, corresponding to $b=1$ in Eq.~\eqref{eq:ell}, and
five different values of $\eta_0$, corresponding to the boundary condition
in Eq.~\eqref{eq:BCalpha} with $n=0,1,2,3,4$. For $\eta_0=0$, red dashed
line, the $\alpha$ field identically vanishes. This case corresponds to the
homogeneous normal phase. With increasing $\eta_0$ the value of $\alpha$ at
the right boundary increases. The values $n=\{1,2,3,4\}$ are respectively
obtained with $\eta_0 \simeq \{ 0.002,0.186, 0.74,1.38 \}$. With reference
to the solid black line, we notice that the $\alpha$ field assumes all the
possible values in the $[0,2\pi]$ interval, while in the homogeneous phase
it can only assume values in the intervals $[0,\pi/2]$ and $[3\pi/2, 2 \pi]$. 
With increasing values of the lattice size the shape of the $\alpha $ field
changes. In Fig.~\ref{fig:alpha_b} we show the plot of the $\alpha$ field
for $n=2$ and three different values of $b$. With increasing values of $b$
the modulation tends to become steeper at $r=\pi$ and flattens at the
boundary. For large values of $b$ the system tends to the homogeneous normal
phase: the integration constant $\eta_0$ decreases with increasing system
size and eventually vanishes for asymptotic values of $b$. We have seen
above that for $\eta_0=0$ one obtains the homogeneous normal phase, however,
in this case we have imposed that $\alpha(2\pi)= 2\pi$, therefore the $%
\alpha $ field discontinuously jumps from $0$ to $2\pi$ at $r=\pi$. The fact
that the large size case corresponds to the homogeneous normal phase can be
seen by combining Eq.~\eqref{eq:ell} and Eq.~\eqref{eq:simplification} in 
\begin{equation}
\frac{\mu_I}{m_\pi} = - \frac{p}{b}\,,
\end{equation}
and therefore asymptotic values of $b$ correspond to vanishing $\mu_I$. 
For numerical evaluations, we shall hereafter assume the values 
\begin{align}
q=1\,, \quad p=1\,, \quad \ell=1/m_\pi \,,\quad \eta_0 \simeq 0.186\,,
\label{eq:values}
\end{align}
where the last equation implies that $\alpha(2\pi) = 2 \pi$.


\subsection{The energy-momentum tensor}

For a given Lagrangian density, $\mathcal{L}$, the energy-momentum tensor is 
\begin{equation}  \label{eq:Tmunu_0}
T_{\mu \nu} = 2 \frac{\partial \mathcal{L}}{\partial g^{\mu \nu}} -g_{\mu
\nu} \mathcal{L}\,,
\end{equation}
and using the expression in Eq.~\eqref{eq:Lgen} we obtain the matter
contribution 
\begin{align}  \label{eq:Tmunu}
T^\text{m}_{\mu \nu } =&-\frac{f_{\pi }^{2}}{2}\mathrm{Tr}\left( \Sigma
_{\mu }\Sigma _{\nu }\right) + \frac{f_{\pi }^{2}}{4} g_{\mu \nu}\mathrm{Tr}
\left( \Sigma^{\alpha }\Sigma _{\alpha }\right)  \notag \\
&- g_{\mu \nu} f_{\pi }^{2} m_\pi^2 \cos\alpha\,.
\end{align}
Upon substituting Eq.~\eqref{eq:general} in Eq.~\eqref{eq:Tmunu} and
normalizing by subtracting the vacuum energy density, see Eq.~%
\eqref{eq:vacuum}, we obtain the matter energy-density 
\begin{align*}
\epsilon_\text{m}= &T^{\text{m}}_{00} + f_{\pi }^{2} m_\pi^2 = \frac{f_\pi^2%
}{2 \ell^2}\biggl( \eta^2+ 2 m_\pi^2 \ell^2(1-\cos{\alpha})  \notag \\
&\left.+ (2p^2\sin^2(q\theta)+q^2)\sin^2\alpha \right. \biggl)\,.
\end{align*}
Taking into account the ranges in Eq.~\eqref{eq:ranges}, the total energy of
the system is 
\begin{equation}
E_\text{m} = \ell^3 \int dr d \theta d \phi \epsilon_\text{m}= \pi
^{2}f_\pi^2\ell\int_{0}^{2\pi }\Omega \left( \alpha \right) d\alpha \ ,
\label{eq:total_energy_m}
\end{equation}
where 
\begin{equation}
\Omega \left( \alpha \right) =\eta (\alpha)+(q^{2}+p^{2})\frac{\sin
^{2}(\alpha )}{\eta (\alpha)} +\ 2 m_\pi^{2}\ell^2 \frac{1-\cos (\alpha )}{%
\eta (\alpha)}\,,  \label{eq:omega}
\end{equation}
depends on the numerical values of the various constants. For the particular
choice in Eq.~\eqref{eq:values} we obtain 
\begin{equation}
E_\text{m} = 20.5 \pi ^{2}\frac{f_\pi^2}{m_\pi}\,.
\end{equation}
The components of the pressure are instead given by 
\begin{align}
P_{rr}\ =& \ \frac{f_{\pi }^{2}}{2\ell ^{2}}(\eta ^{2}-q^{2}\sin ^{2}{\alpha 
}+2m_{\pi }^{2}\ell ^{2}\cos {\alpha })\,,  \notag \\
P_{\theta \theta }\ =& \ \frac{f_{\pi }^{2}}{2\ell ^{2}}(-\eta
^{2}+q^{2}\sin ^{2}{\alpha }+2m_{\pi }^{2}\ell ^{2}\cos {\alpha })\,,  \notag
\\
P_{\phi \phi }\ =& \ \frac{f_{\pi }^{2}}{2\ell ^{2}}\left( 2\ell ^{2}m_{\pi
}^{2}\cos {\alpha }-\eta ^{2}-q^{2}\sin ^{2}{\alpha }\right.  \notag \\
& \left. +2\sin ^{2}(\alpha )\sin ^{2}(q\theta )p^{2}\right) \,,
\label{eq:Ps}
\end{align}
showing that the pressure is not isotropic. This happens because of the
space modulation of the $\alpha$ and $\Theta$ fields. When substituting Eq.~%
\eqref{eq:eta} in Eqs.~\eqref{eq:Ps} we obtain 
\begin{align}
P_{rr}\ =& \frac{f_{\pi }^{2}}{2\ell ^{2}}(\eta _{0}^{2}+2m_{\pi }^{2}\ell
^{2})\simeq f_{\pi }^{2}m_{\pi }^{2}\,,  \notag \\
P_{\theta \theta }\ =& \frac{f_{\pi }^{2}}{2\ell ^{2}}(2m_{\pi }^{2}\ell
^{2}(2\cos \alpha -1)-\eta _{0}^{2})\simeq f_{\pi }^{2}m_{\pi }^{2}(2\cos
\alpha -1.01) \ ,  \notag \\
P_{\phi \phi }\ =& \frac{f_{\pi }^{2}}{2\ell ^{2}}(2m_{\pi }^{2}\ell
^{2}(2\cos \alpha -1)+2\sin ^{2}(\alpha )(p^{2}\sin ^{2}(q\theta )-q^{2}) \ ,
\notag \\
& -\eta _{0}^{2})\simeq f_{\pi }^{2}m_{\pi }^{2}(2\cos \alpha +\sin
^{2}(\alpha )(\sin ^{2}(\theta )-1)-1.01) \ ,
\end{align}
where the last equalities are obtained using Eq.~\eqref{eq:values}. The fact
that the pressure in a certain region of the $(\theta ,\phi )$ plane becomes
negative follows from the fact that the system is not static, but
stationary, therefore cavitation-like phenomena are possible. Finally, we
note that 
\begin{equation}
T_{0\phi }=-\frac{f_{\pi }^{2}}{\ell ^{2}}\sin ^{2}{\alpha }\sin
^{2}(q\theta )p^{2}=-f_{\pi }^{2}m_{\pi }^{2}\sin ^{2}{\alpha }\sin
^{2}(\theta )\,,  \label{eq:T0phi}
\end{equation}
where the last equation holds for the values in Eq.~\eqref{eq:values}. The
presence of a persistent current means that there is a continuous steady
energy transfer in the $\phi $ direction, which is due to the presence of a
stationary flow.


\subsection{Topological charge}

\label{sec:radial_topo} 

\begin{figure*}[thb!]
\includegraphics[width=0.4\textwidth]{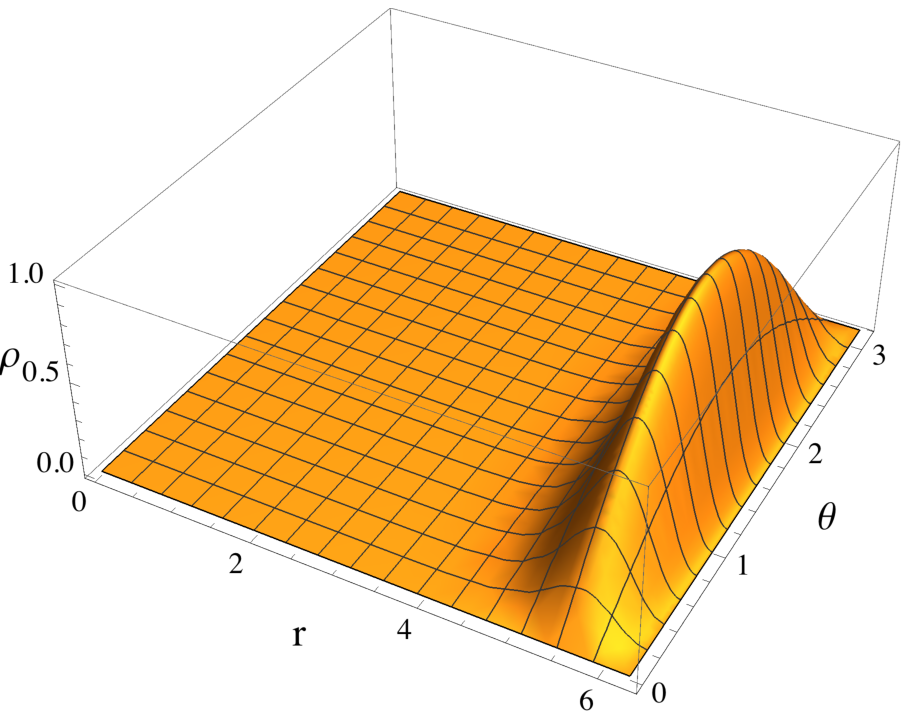} \hspace{1.cm} %
\includegraphics[width=0.4\textwidth]{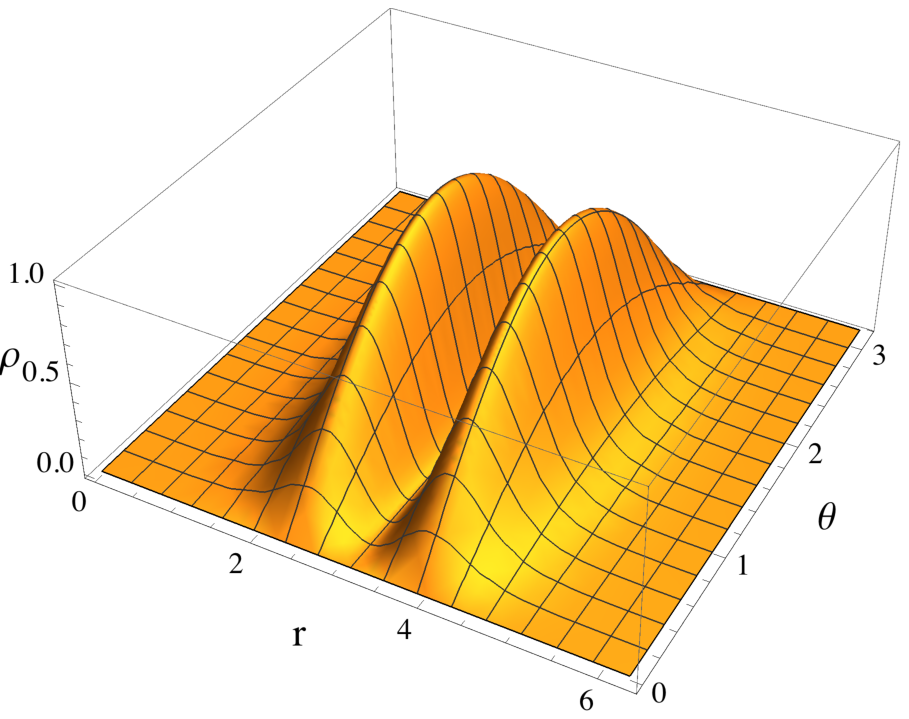}
\caption{Modulation of the topological density contribution of the matter
fields, see Eq.~\eqref{eq:rhom2}, obtained for $n=1$, left hand side, and $%
n=2$, right hand side. The topological density has been normalized to the
value at the maximum. }
\label{fig:solitons}
\end{figure*}
The topological charge of the solitonic configuration can be written as 
\begin{equation}
B_\text{m}=\frac{\ell^3}{24\pi ^{2}}\int drd\theta d\phi \ \rho _{\text{m}}\
\,,  \label{eq:topo1}
\end{equation}%
where 
\begin{equation}  \label{eq:rhom}
\rho _{\text{m}}=\epsilon ^{ijk}\text{Tr}\left\{ \left( \Sigma ^{-1}\partial
_{i}\Sigma \right) \left( \Sigma ^{-1}\partial _{j}\Sigma \right) \left(
\Sigma ^{-1}\partial _{k}\Sigma \right) \right\} \ ,
\end{equation}%
is the topological density contribution of the matter fields. The expression
in Eq.~\eqref{eq:rhom} shows that when the $SU(2)$-valued scalar field $%
\Sigma $ depends on one or two coordinates, the topological density
identically vanishes. Upon substituting Eq.~\eqref{eq:Sigma} in Eq.~%
\eqref{eq:rhom}, the topological density can be rewritten as 
\begin{align}
\rho_{\text{m}} =-12 \frac{pq}{\ell^3}\sin (q\theta )\sin ^{2}(\alpha ) \eta
\label{eq:rhom2}
\end{align}
which can be readily integrating noticing that 
\begin{equation}
\int dr \sin ^{2}(\alpha) \eta = \int d\alpha \sin^2 \alpha\,,
\label{eq:alpha_int}
\end{equation}
and since we have imposed the boundary condition in Eq.~\eqref{eq:BCalpha},
the topological charge becomes 
\begin{equation}
B_{\text{m}}=%
\begin{cases}
-n p & \text{if}\quad q\text{ odd} \ , \\[1.5ex] 
0 & \text{if}\quad q\text{ even} \ ,%
\end{cases}
\label{eq:topological_charge}
\end{equation}%
which clarifies the choice of the $q$ parameter in Eq.~\eqref{eq:general}.
We report in Fig.~\ref{fig:solitons} the topological charge densities for
the cases with $n=1$, left panel, and $n=2$, right panel. Physically, the
topological charge represents the baryonic charge of the system~\cite%
{Callan:1983nx,Piette:1997ny}. Therefore, the spatial modulation of the
fields is associated with the realization of a non-vanishing baryonic
density. With reference to the $q=1$ and $n=2$ case we see from Eqs.~%
\eqref{eq:rhom2} and \eqref{eq:alpha_int} that the two maxima of the
topological charge density correspond to $\theta=\pi/2$ and $\alpha = \pi/2$
and $\alpha = 3\pi/2$. The normal and condensed phases correspond to values
of $\alpha$ in the intervals $[0,\pi/2]$ and $[3\pi/2,2\pi]$, see Eq.~%
\eqref{eq:alpha_hom}. These facts suggest that the spatial modulation of the 
$\alpha$ field in the interval $[0,2\pi]$ describes two (anti)baryons,
approximately realized in the interval $[\pi/2,3\pi/2]$, surrounded by a
cloud of pions forming a condensate that vanishes at the boundary of the
system and reaching its maxima at the places where the the baryonic density
takes its maximal values. Systems with a larger baryonic charge correspond
to larger values of $n$.


\section{Inhomogeneous phase including the gauge field}

\label{sec:gauge} 
By including the electromagnetic interaction the matter Lagrangian turns to 
\begin{align}  \label{eq:Lgen_2}
\mathcal{L}_\text{m} = & \frac{f_\pi^2}2 \left[ \partial_\mu \alpha
\partial^\mu \alpha +\sin^2{\alpha} \partial_\mu \Theta \partial^\mu \Theta
+ 2 m_\pi^2 \cos\alpha \right.  \notag \\
&\left. +\sin^2{\alpha} \sin^2{\Theta}(\partial_\mu \Phi-2 \tilde{A}%
_\mu)(\partial^\mu \Phi-2 \tilde{A}^\mu) \right]\,,
\end{align}
where $\tilde{A}^\mu$ is defined in Eq.~\eqref{eq:deftilda}. From this
expression, it is clear that only the $\Phi$ field is minimally coupled to
the electromagnetic field. In the following we shall work in the Lorenz
gauge $\partial_\mu A^\mu =0$; by using the particular classical fields in
Eq.~\eqref{eq:general} and 
\begin{equation}  \label{eq:au}
A^\mu =(u,0,0,u)\,,
\end{equation}
the $\alpha(r) $ field decouples and its classical solution satisfies Eq.~%
\eqref{eq:alpha_sol}. This means that although the gauge field is generated
by the pions, for the particular choice in Eq.~\eqref{eq:au} it does not
back react on the classical fields.

\begin{figure}[h!]
\includegraphics[width=0.45\textwidth]{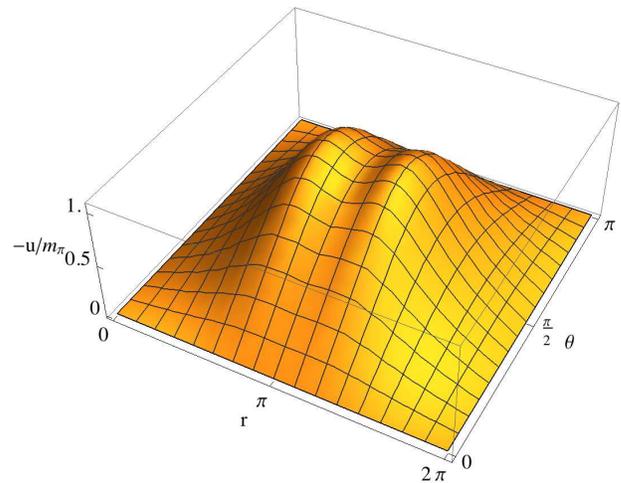}
\caption{Modulation of the electromagnetic potential as a function of the $r$
and $\protect\theta$ coordinates obtained solving the differential equation 
\eqref{eq:maxred3}, where $\Psi=1/\ell-2 u$, assuming that $u=0$ at the
boundary. }
\label{fig:un}
\end{figure}

Let us now comment on the particular ansatz in Eq.~\eqref{eq:au} for the
electromagnetic potential. It does not correspond to a particular gauge, but
to a particular configuration of the electric and magnetic fields. With this
ansatz we have that 
\begin{align}
\bm E= -\frac{1}{\ell}(\partial_r u, \partial_\theta u, 0)\,, \qquad \bm B= 
\frac{1}{\ell}(\partial_\theta u, - \partial_r u, 0)\,,
\end{align}
thus the electric and magnetic fields have equal magnitude $|\bm E|^2= |\bm %
B|^2= ((\partial_r u)^2 + (\partial_\theta u)^2)/\ell^2 $ and they are in
the $r-\theta$ plane. It remains to be determined the expression of $u$
using the Maxwell equations. From Eq.~\eqref{maxcurrent} or Eq.~%
\eqref{eq:Lgen} we obtain the electromagnetic current 
\begin{align}  \label{eq:Jmu}
J_\mu = - 2f_\pi^2 \sin^2{\alpha} \sin^2 (q \theta) (\partial_\mu \Phi - 2 
\tilde{A}_\mu) \,,
\end{align}
therefore the non-vanishing components of the current are 
\begin{align}
J_0 & =-2 \frac{f_\pi^2}{\ell} \sin ^{2}(\alpha )\sin ^{2}(q\theta )\left(
p-2 \ell u\right) \ ,  \label{eq:curr1} \\
J_{\phi }& =-J_0 \,,  \label{eq:curr2}
\end{align}%
where we used Eqs.~\eqref{eq:ell} and \eqref{eq:simplification} with $b=1$,
hence Eq.~\eqref{eq:maxwellNLSM} reduces to the one single equation 
\begin{align}
\frac{1}{\ell^2}\left(\frac{\partial^2 }{\partial r^2} + \frac{\partial^2 }{%
\partial \theta^2}\right)u = 2 \frac{f_\pi^2}{\ell} \sin^2{\alpha}%
\sin^2\Theta \left(p-2 \ell u\right) \,,
\end{align}
where we have used Eq.~\eqref{eq:au} and the Lorenz gauge condition. This
expression can be rewritten as a time-independent Schr\"odinger-like
equation in a periodic two-dimensional potential 
\begin{equation}
\left( \frac{\partial ^{2}}{\partial r^{2}}+\frac{\partial ^{2}}{\partial
\theta ^{2}}\right) \Psi +V\Psi =0\ ,  \label{eq:maxred3}
\end{equation}%
where 
\begin{equation}
\Psi =\frac{p}{\ell}-2u \quad \text{and} \quad V =4 f_\pi^2 \ell^2
\sin^2\alpha \sin^2(q \theta)\, ,  \label{eq:transf}
\end{equation}
are the wave function and the effective potential, respectively. As boundary
condition we assume the simplest one, that is $u=0$ along the whole
boundary. The resulting plot of the potential is in Fig.~\ref{fig:un}, where
we have taken $q=1$ and $p=1$. We tried different boundary conditions with a
constant shift of the potential at different boundaries, obtaining similar
results.

\begin{figure*}[t!]
\includegraphics[width=0.4\textwidth]{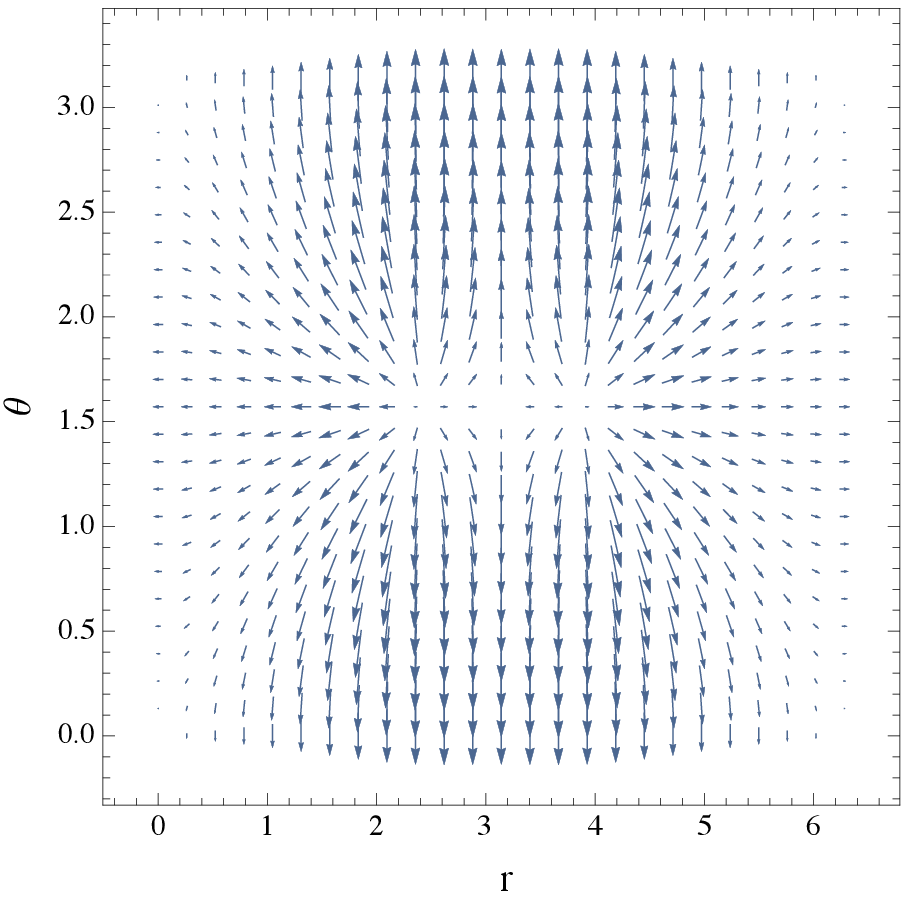} \hspace{1cm} %
\includegraphics[width=0.4\textwidth]{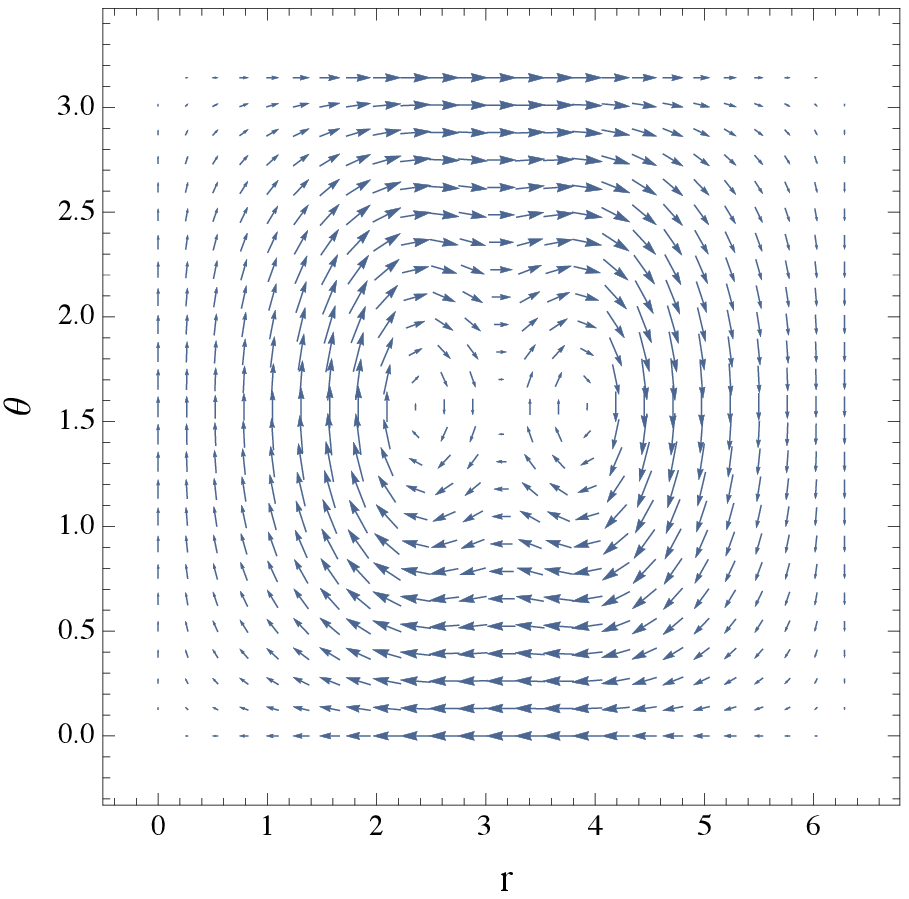}
\caption{Electric field (left) and magnetic field (right) obtained with the
values in Eq. \eqref{eq:values} and assuming that the electromagnetic
potential vanishes at the boundary.}
\label{fig:emfields}
\end{figure*}

\begin{figure}[h!]
\includegraphics[width=0.45\textwidth]{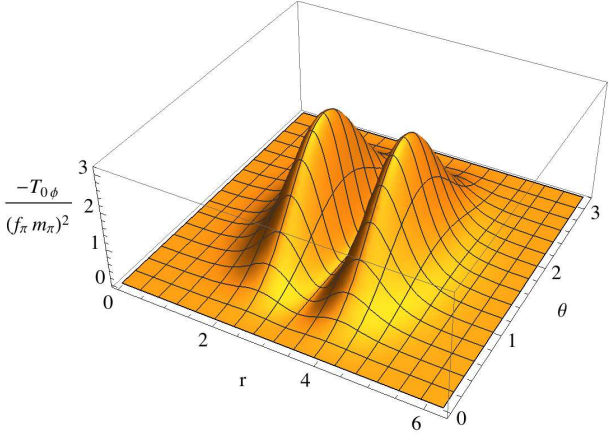}
\caption{Intensity of the current $T_{0\protect\phi}$ defined in Eq.~ 
\eqref{eq:T0phi} flowing along the $\protect\phi$ direction orthogonal to
the $r-\protect\theta$ plane. The current is concentrated in the regions
where the magnetic field reaches its minimum and the topological matter
density as well the energy density reach their maximum values. }
\label{fig:current}
\end{figure}

The corresponding electric and magnetic fields are reported in Fig.~\ref%
{fig:emfields}. The electric field is centered at the maxima of the two
solitons and decreases its intensity approaching the soliton centers. This
shows that the electric field is screened inside the solitons. Similarly,
the magnetic field in the right panel of Fig.~\ref{fig:emfields} is screened
inside the solitons, where the electromagnetic current reaches its maximum,
see Fig.~\ref{fig:current}. Therefore, there is a form of Meissner screening
induced by the persistent electromagnetic current $J^\phi$ flowing
perpendicularly to the $r-\theta$ plane.

Summarizing, we have reduced the three coupled field equations~in Eq. %
\eqref{eq:NLSM} and the four coupled Maxwell equations in Eq. %
\eqref{eq:maxwellNLSM} to the two differential equations in Eqs.~%
\eqref{eq:alpha_sol} and \eqref{eq:maxred3}. This has been accomplished by
the particular ansatz for the angular classical fields in Eq.~%
\eqref{eq:general} and the particularly choice of the gauge field in Eq.~%
\eqref{eq:au}. This ansatz greatly simplifies the problem. The linear
response of the system to the gauge field, Eq.~\eqref{eq:Jmu}, shows that
the produced current has the same form of the energy density current in Eq.~%
\eqref{eq:T0phi}, with the additional gauge field contribution.

Indeed, in the presence of the electromagnetic interaction, there are two
additional contributions to the energy-momentum tensor, one is a
contribution determined by the minimal coupling with the $\Phi $ field and
the other is the pure gauge one. The minimal coupling can be taken into
account using Eq.~\eqref{eq:Tmunu}, where now $\Sigma _{\mu }$ is defined in
Eq.~\eqref{eq:sigmamu} and includes the gauge field. The pure gauge
contribution has the standard expression 
\begin{equation}
T_{\mu \nu }^{\text{em}}=-F_{\mu \alpha }F_{\nu }^{\;\alpha }+\frac{1}{4}%
F_{\alpha \beta }F^{\alpha \beta }g_{\mu \nu }\,,  \label{eq:tmunu}
\end{equation}%
which turns in 
\begin{equation}
T_{\mu \nu }^{\text{em}}=|\bm E|^{2}\left( 
\begin{array}{cccc}
1 & 0 & 0 & 1 \\ 
0 & 0 & 0 & 0 \\ 
0 & 0 & 0 & 0 \\ 
1 & 0 & 0 & 1%
\end{array}%
\right) \,,
\end{equation}%
using the particular expression in Eq.~\eqref{eq:au} of the gauge potential.
The energy-density now turns in 
\begin{align}
\epsilon =& \frac{f_{\pi }^{2}}{2\ell ^{2}}\biggl( \eta ^{2}+
[q^{2}+2(p-2\ell u)^2\sin ^{2}(q\theta )]\sin ^{2}{\alpha }  \notag \\
& \ \ \ + 2m_{\pi }^{2}\ell ^{2}(1-\cos {\alpha })\biggl) +|\bm E|^{2}\,,
\end{align}%
and taking into account the ranges in Eq.~\eqref{eq:ranges}, the total
energy of the system is 
\begin{equation}
E=E_{\text{m}}\ +E_{1}+E_{\text{g}}\,,  \label{eq:total_energy_m2}
\end{equation}%
where $E_{\text{m}}$ is in Eq.~\eqref{eq:total_energy_m} and 
\begin{equation}
E_{1}\simeq 17.5\frac{f_{\pi }^{2}}{m_{\pi }}\,,\quad E_{\text{g}}\ \simeq
2.2m_{\pi }\,,
\end{equation}%
are obtained considering $\ell =1/m_{\pi }$.

The $P_{rr}$ and $P_{\theta\theta}$ pressure components have the same
expressions reported in Eq.~\eqref{eq:Ps}, while 
\begin{align}
P_{\phi\phi} \ =& \ \frac{ f_\pi^2}{2 \ell^2} \left( 2 \ell^2 m_\pi^2 \cos{%
\alpha} -\eta^2 - q^2 \sin^2{\alpha} \right.  \notag \\
& \left.+2 \sin^2(q\theta) \sin^2(q\theta)(-p+2 \ell u )^2\right) + |\bm %
E|^{2}\,.
\end{align}
Thus there are additional gauge field contributions only to the pressure in
the $\phi$ direction. As before, the energy density and the pressure are not
time dependent: the system is stationary with a constant energy transfer in
the $\phi$ direction given by 
\begin{equation}
T_{0 \phi} = -\frac{f_\pi^2}{\ell} \sin^2{\alpha}\sin^2(q\theta)(p-2\ell
u)^2 - |\bm E|^2 \ ,
\end{equation}
where the last term on the right hand side is the electromagnetic Poynting
vector.

For completeness, we notice that when gauging the $U(1)$ symmetry, one
should include an additional contribution to the topological charge~\cite%
{Callan:1983nx,Piette:1997ny}, which now reads 
\begin{equation}
B=\frac{\ell^3}{24\pi ^{2}}\int drd\theta d\phi \ (\rho _{\text{m}}+\rho
_{g})=B_{\text{m}}\ +B_{\text{g}}\ ,  \label{eq:topo2}
\end{equation}%
where $B_{\text{m}}$ is defined in Eq.~\eqref{eq:topo1}. The gauge field
contribution 
\begin{equation}  \label{eq:rhog}
\rho _{g}=-i \epsilon ^{ijk}\text{Tr}\left\{ \partial _{i}\left[ 3
A_{j}\sigma_{3}\left( \Sigma ^{-1}\partial _{k}\Sigma +\left( \partial
_{k}\Sigma \right) \Sigma ^{-1}\right) \right] \right\} \,,
\end{equation}%
is the Callan-Witten topological charge: it guarantees both the conservation
and the gauge invariance of the topological charge. For the considered
potential $A_{j} = \delta_{j3}u$, hence this topological density can be
written as 
\begin{equation}
\rho _{g}= - 3 i \epsilon^{i3k}\partial_i u T_k= \frac{3 i}{\ell}%
[\partial_r(u T_2) -\partial_\theta(u T_1) ]\,,
\end{equation}
where 
\begin{equation}
T_k= \text{Tr} \left[\sigma_{3}\left( \Sigma ^{-1}\partial _{k}\Sigma
+\left( \partial _{k}\Sigma \right) \Sigma ^{-1}\right) \right]\,,
\end{equation}
determine the dependence on the solitonic field. Then, for the considered
solitonic configuration we have that 
\begin{equation}
B_g = \frac{i \ell^2}{4 \pi} \left\{\int_0^\pi d \theta \left. u
T_2\right\vert_{r=0}^{r=2 \pi} - \int_0^{2 \pi} d r \left. u
T_1\right\vert_{\theta=0}^{\theta= \pi} \right\}\,,
\end{equation}
where 
\begin{equation}
T_1 = 4 i \frac{\eta}{\ell} \cos\Theta \ , \qquad T_2= -2 i \frac{q}{\ell}
\sin 2\alpha \sin\Theta\,.
\end{equation}
Since along the boundary the gauge field takes the fixed value $\bar u$ and
since $\sin2\alpha(r=2 \pi) = \sin2\alpha(r=0)=0$, see Eq.~\eqref{eq:BCalpha}%
, we readily have that $\left. u T_2\right\vert_{r=0}^{r=2 \pi} =0$. Then 
\begin{equation}
B_g = \frac{\ell}{\pi}\bar u ( \cos(q \pi) -1)\int_{0}^{2 \pi} dr \eta\,.
\end{equation}
For $\bar u=0$, corresponding to the boundary condition used to obtain Fig.~%
\ref{fig:alpha_b}, we have that $B_g=0$, meaning that there is no
contribution of the gauge fields to the topological charge.

For different boundary conditions the topological charge may give a
non-vanishing contribution. For instance, for $\bar u =p/2 \ell$,
corresponding to vanishing currents at the boundary, we have that 
\begin{equation}
B_g=%
\begin{cases}
-n p & \text{if}\quad q\text{ odd} \ , \\[1.5ex] 
0 & \text{if}\quad q\text{ even} \ ,%
\end{cases}
\label{eq:topological_charge_CW}
\end{equation}%
and comparing with Eq.~\eqref{eq:topological_charge}, we find that $%
B=B_m+B_g = 2 B_m = - 2n p$, where the last equality holds for $q$ odd.


\section{Conclusions and Perspectives}

\label{sec:conclusions} 

We have derived the first analytic example of topologically non-trivial
inhomogeneous pion condensates in low-energy QCD in (3+1)-dimensions at
finite isospin chemical potential. We have shown that for a particular
condensate ansatz, the complete set of field equations both with and without
the minimal coupling to the electromagnetic field can be consistently
reduced to integrable equations, where the isospin chemical potential can be
related to the system size. In these configurations the topological charge
does not vanish and it is deeply related to the stationary properties of the
system. We find that the energy density $\epsilon_\text{m}$, the topological
density $\rho_\text{m}$ and the currents $J_{0}$, $J_{\phi }$ and $T_{0\phi}$
are constant in the $\phi $ direction while they depend non-trivially on the
two spatial coordinates $r$ and $\theta $. The regions of maximal $\rho_%
\text{m}$ are three-dimensional tubes of length $2\pi \ell$ parallel to the $%
\phi $ direction (the same is true for $\epsilon$ and for the currents),
forming a pasta-like phase (see \cite{Ravenhall:1983uh,10.1143/PTP.71.320,
Horowitz:2014xca} and references therein for pasta phases in nuclear
physics). It is worth emphasizing that the proposed inhomogeneous phase
avoids the Derrick's scaling argument for the existence of solitonic
solutions because it is realized in a finite volume with appropriate
non-trivial boundary conditions. Moreover, it is not static, because the $%
J_{0}$, $J_{\phi }$ and $T_{0\phi}$ currents do not vanish even when the
electromagnetic potential vanishes ($u=0$), and are maximal where $\sin
^{2}(\alpha )\sin ^{2}(q\theta )=1$, see Eqs.~\eqref{eq:T0phi} and %
\eqref{eq:curr1}, rapidly decreasing far from the peaks. 
Consequently, these currents cannot be turned off continuously: they are
topologically protected. Therefore, these stationary currents are actually
supercurrents. If the broken $U(1)$ symmetry is global, they correspond to
superfluid currents. If the broken $U(1)$ symmetry is gauged, then they
correspond to superconducting currents. Indeed, the typical expression of
any supercurrent, see for instance~\cite{Witten:1984eb}, is 
\begin{equation}
\bm{J} \sim \kappa \left( \bm \nabla \varphi +{\bm A}\right) \ ,
\label{eq:wittsc2}
\end{equation}%
where $\kappa$ is a constant, $\varphi $ is a phase and ${\bm A}$\ is the $%
U(1)$ gauge potential. In the standard settings \cite{Witten:1984eb} there
is no topological number associated with $\kappa $: the linear stability of
the configurations where $\kappa\neq 0$ is established by direct methods
(such as linear perturbation theory) and is determined by a control
parameter. In the present case, the factor $\sin ^{2}(\alpha )\sin
^{2}(q\theta )$ in the currents (which plays the role of $\kappa$ in Eq.~%
\eqref{eq:wittsc2}) is instead topologically protected.

As far as we know, these are the first analytic examples of gauged
crystal-like structures in low-energy QCD. The realized configuration can be
interpreted as consisting of $n$ baryons embedded in an inhomogeneous pion
gas, characterized by the modulation of the $\alpha$, $\Theta$ and $\Phi$
fields. It would be interesting to compare the inhomogeneous and the
homogeneous condensates in order to establish which one is more convenient
thermodynamically. At a first glance, we could do this explicitly since we
can compute the free energy (in a grand canonical ensemble with a chemical
potential $\mu_T$ associated with the topological charge) both for the
homogeneous and for the inhomogeneous condensates. However, the two
condensates seem to be realized in different regimes. If we identify the
topological charge with the baryonic charge, the inhomogeneous system
corresponds to baryons embedded in a pion gas forming a condensate that
reaches its maximum values where the baryonic density is larger. On the
other hand, the homogeneous phases {\ attainable within our framework}
correspond to pure pionic systems. Therefore the two systems correspond to
two different grand canonical ensembles. In principle, it is possible to
extend chiral perturbation theory including baryons, see for instance~\cite%
{Pich:1998xt,Scherer:2002tk, Scherer:2005ri}, however our procedure is
different, because we have introduced baryons as topological objects in a
cloud of pions with a non-vanishing isospin asymmetry.

Remarkably, in the presence of a non-vanishing isospin chemical potential it
is possible to realize the solitonic configuration eliminating any time
dependence of the condensate, thus the boundary conditions are time
independent, as well. This is a clear improvement with respect to the case
of vanishing $\mu_I$, in which the solitonic configuration has been realized
with a time dependent $\Phi$ field and then with time dependent boundary
conditions~\cite{Canfora:2013osa, Canfora:2013xja, PhysRevD.89.025007,
Canfora:2014aia, Canfora:2015xra, Ayon-Beato:2015eca, Tallarita:2017bks,
Canfora:2017ivv, Giacomini:2017xno, Astorino:2018dtr, Alvarez:2017cjm,
Aviles:2017hro, Canfora:2018clt, Canfora:2018rdz,
Canfora:2019asc,Canfora:2019kzd, Alvarez:2020zui, Canfora:2020kyj,
Canfora:2020ppn}. We have found that when the isospin asymmetry is related
to the system size by Eq.~\eqref{eq:simplification} any time dependence is
eliminated, and only in this case the topologically stable crystalline phase
is time independent. It would be interesting to compare our results with
those of LQCD simulations, which should realize the condition in Eq.~%
\eqref{eq:simplification} by properly modulating the ratio between the
lattice size and the isospin chemical potential. It may well be that the
LQCD simulations find that a different configuration is energetically
favored. Although stable, the solution we obtained may not correspond to the
energetically favored solitonic configuration: to derive our results we have
indeed employed a number of simplifying assumptions.

We also notice that in principle one can use numerical techniques to solve
the set of differential Eqs.~\eqref{eq:Phi}, \eqref{eq:Theta} and %
\eqref{eq:alpha}. The comparison with the outcome of these numerical
simulations would be quite interesting, as well. In any case, as we have
shown, the analytic solutions are not just of academic interest as they
disclose relevant physical properties of complex structures that may somehow
be hidden in the numerical procedures.

Given the non-trivial crystalline structure of our system, it would also be
interesting to investigate the spectrum of its low-energy excitations. While
our configurations are protected by their topological charge against decay,
the presence of non-trivial pole structures at finite momentum might signal
strong fluctuation effects (see eg. \cite{Pisarski:2020dnx} for a discussion
on how an inhomogeneous chiral phase is turned into a quantum spin liquid by
fluctuations). For the modulations considered in this work, we expect at
least some of the low-energy excitations to have a phonon-like linear
dispersion relation (see \cite{Takayama:1992eu} for a study in a similar
scenario), but a more detailed investigation would definitely be of
interest, and we plan to come back to this issue in a future work.

We note that we have derived our results employing leading order $%
\chi $PT, however the results reported in~\cite{Alvarez:2020zui,
Canfora:2020kyj, Canfora:2020ppn}, where additional interaction terms have
been introduced, strongly suggest that the present construction could be
valid even beyond leading order $\chi $PT.

{If the crystals discussed in this work are realized in dense stellar objects, they may have various phenomenological effects on their properties. Since a solitonic crystal supports superfluid and/or superconductive flows, it would manifest in a pasta-like structure that  influences the thermal transport properties inside the star.
Moreover, it should be characterized by a certain rigidity, meaning that deforming the structure should result in a certain energy cost. Thus, for rotating compact stars, it  could be relevant for vortex pinning and may be associated with peculiar stellar glitches. 
Finally, our results suggest that even   pion stars, completely made by pions and electrons \cite{Carignano:2016lxe,Brandt:2018bwq,Andersen:2018nzq}, could have a rigid crust made by a solitonic crystal.  In this case, this phase should manifest not only in peculiar  stellar glitches, but it could as well  support quadrupolar mass deformations associated with the continuous emission  of gravitational waves.   We plan to come back on these very interesting issues in a future publication.}

\subsection*{Acknowledgements}

We thank Rob Pisarski for useful discussions. F. C. has been funded by
Fondecyt Grants 1200022. M. L. and A. V. are funded by FONDECYT
post-doctoral grant 3190873 and 3200884. The Centro de Estudios Cient\'{\i}ficos
(CECs) is funded by the Chilean Government through the Centers of Excellence
Base Financing Program of Conicyt. S. C. has been supported by the MINECO
(Spain) under the projects FPA2016-76005-C2-1-P and PID2019- 105614GB-C21,
and by the 2017-SGR-929 grant (Catalonia).

\bibliography{BIB}

\end{document}